# Multiport Rapid Charging Power Converter

Yikun Chen
*School of Electronics, Electrical Engineering and Computer Science (EEECS), Queen's University Belfast*
*Belfast, BT9 5AH, UK*
ychen47@qub.ac.uk

Ahmad Elkhateb
*School of Electronics, Electrical Engineering and Computer Science (EEECS), Queen's University Belfast*
*Belfast, BT9 5AH, UK*
A.Elkhateb@qub.ac.uk

*Abstract*—Rapid charger is getting more and more important as the electric vehicle (EV) getting popular. The rapid charging technique plays an important part in the electric vehicle development. Multiport converter is used in the rapid charging technique to reduce the required current and also provides some other advantages.

In this paper, the literature review form multiport converter to the rapid charger of electric vehicle is introduced. Some topologies of multiport converter are discussed in the literature review. Triple active bridge topology (TAB) is selected as it is useful and commonly used. The feasibility of using Sinusoidal Pulse Width Modulation (SPWM) to control the converter is also discussed in the simulation part. A simple hardware design and experiment is included at the end.

*Keywords*—Multiport converter, *rapid charging, pulse width modulation.*

## I. INTRODUCTION

As the industries develop, the conventional two port converter cannot meet all the needs of the industries. Fields such as solar cell array, wind turbines, fuel cells etc. need a multiport converter instead of a two-port converter or the traditional converters to facilitate low ripple operation and enhance paralleling of the input side **[1-6]**. Multiport converter shows some good features such as ease of matching different voltage levels at the output port, more compact dimensions of the design and installation, possibility of achieving a better efficiency and lower cost, an increased dynamical performance and management due to the centralized control, provide the storage port to manage the power flow and so on [7]. For instance, a solar system needs storage batteries port for storing excess power and resupplying the stored power to the load when it is needed [8-10]. (Such as the rainy day or during the night time.)

The multiport converter can be separated into three categories by how it connects each port as shown in Fig. 1. Fig. 1 (a) shows the multiport converter with DC link, such as the topology in [11] and [12]. Fig. 1 (b) shows the multiport converter that contains separating winding for each port, such as the topology in [16-15]. Fig. 1 (c) shows the multiport converter combining DC link and magnetic winding, which allows some ports share a common ground, such as the topology in [16-18]. As the magnetic windings provide an extra isolation among the ports, Fig. 1 (b) has a better isolation than Fig. 1 (b) and Fig. 1 (b) has a better isolation than Fig. 1 (a).

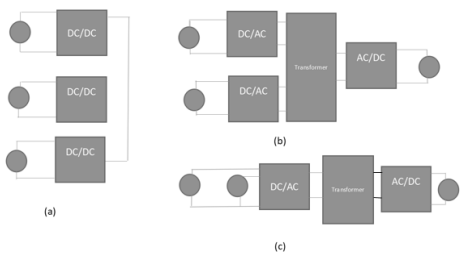

Fig 1. Three categories of multiport converter

Among various multiport converter topology, there is one kind of topology have some attractive characteristics, which is called three port tripe-active-bridge (TAB) topology. It is driven from dual-active-bridge (DAB) topology. As early as 1991, DAB topology has been extended to TAB topology in [19] as it shows some attractive features such as low device stresses, bidirectional power flow, ease of power control, good isolation among each port etc. Then this topology is proposed in many paper such as [20-27].

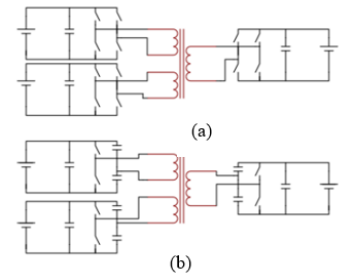

Fig 2. Two types of TAB topology

This kind of topology uses active bridges for each port and transformer to link each port together. For the input port, the function of the active bridges is to change the DC input to AC for transferring the power through the transformer. The function of the active bridges of the output port is to change the current transferred from the transformer to DC (act as a rectifier). We can control the power flow or the power direction simply through the active bridges. Fig. 2 shows two types of TAB topology. Fig. 2 (a) is the full bridge topology and Fig. 2 (b) is the half bridge topology.

In [20] and [31], both papers present a three-port bidirectional converter for hybrid fuel cell systems that improve the slow transient response of the PEM fuel cell stack. In [22], a three-phase power supply systems formed by TAB topology is presented. Duty ratio control is used for adjusting the amplitude of the fundamental component. [20] [21] and [22] are all not extended the zero-voltage-switching range (ZVS). Zero-voltage-switching is a soft switching technique. This technique makes the voltage across the switch is zero when switching. It is very important when the switch is operated with a high frequency due to the losses of the switching will become non-negligible, which is very common in the TAB topology. However, many TAB converter such as [20-22] cannot operate with zero-voltage-switching over the full operation region with wide input ranges. To solve this problem, [23] [24] come out with the idea of adding parallel inductors and voltage cancellation. However, the method is complex and difficult to implement. Then [25] is proposed to solve the problem by a simplified duty cycle control method, which is simple, effective and easy to implement. Also, the soft start-up methods were discussed. Furthermore, soft-switching over full operation range are achieved in [26] and [27] by adjusting the duty cycle continuously.

To realize the control of the multiport converter, pulse width modulation (PWM) technique is commonly used. With pulse width modulation, we can simply control the power flow by adjusting the duty cycle. Also phase shifting technique is used for realizing the bidirectional function and soft switching.

Although there are various multiport converters being proposed. However, different industry has different requirements in such as voltage range, current rating, size, cost, etc. Many of the proposed multiport converter topologies meet some shortcoming, which means that they are not suitable in some industries.

In [28], a method based on a time share concept is proposed, but it only allows one port to transfer energy from or to the common magnetic winding at one time, which means that the power rating is limited. Also, the converter presented in [29] is not suitable for high-power application as it is a half-bridge topology that has limitations of power rating. (Each bridge needs to tolerate the same voltage of the input.) And the converter in [30] bases on flux additivity in a multi-winding transformer, with capability of simultaneous power transfer. However, its reverse-blocking diodes only allow unidirectional power flow, so we cannot use it in some application relating to energy storage. [31]

The converter in [32] is suitable for renewable energy systems as it has the bulk energy storage. And the converter in [33] is suitable for medium-power applications where simple topology, autonomous operation, compact design and installation and low cost are required.

One problem that EV is facing is that, unlike the conventional fuel vehicle that can be refilled directly, EV takes time to recharge. As the TABLE 2 shown, it may take hours, which is too long for the customers to wait. So, the rapid charging technique for electric vehicle is in need. To charge rapidly, high current is needed, which means that the charging system should have high current ratings as well as the ability to provide the high current. So, a storage port is needed, and that is the reason why we use multiport converter for the charger. Generally, three-port converter is needed, as one port for the grid, one port for the storage battery and one port for the EV battery. The storage battery can be charged during the hours of off-peak grid demand time. Then EV battery can be charged by the storage battery and grid at the same time, which reduce the required current from grid and therefore, the impact on the grid and the power rating of the converter will be reduced. In addition, whether the port is bidirectional is taken to consideration, as the storage can also support the grid when it is needed.

[37], [22] and [38] use the tripe-active-bridge (TAB) topology with high frequency transformer to realize rapid charging.

In [37], matrix converter is implemented on the grid port, three-leg inverter is used on vehicle battery port and six-leg inverter is used on storage battery port. The six-leg inverter gives some advantages such as lower conduction losses, decreasing the size of output filter and input capacitor, zero voltage zero current switching over a wide operation range without other extra components. And matrix converter has high efficiency and provides convenience in transformer design. The main drawback of this multiport converter is that 42 switches are used, which causes the complexity of the design and control.

[22] proposes a three-phase three-port high-frequency link converter, which consists of multiple four-quadrant full bridge cells. Controllability of the power flow of this converter is discussed. This converter only uses 24 switches, so it is easier in design and more reliable than [37]

In [38], the proposed multiport converter consists of six-leg inverter for the storage port, cycloconverter for the grid port, three-leg inverter for the vehicle battery port and the high-frequency link transformer to link each port. The design and the simulation test are included in the paper. 24 switches are used.

So, tripe-active bridge topology is selected as the desired topology for this project. It contains three bridges and a high frequency transformer.

As the flux density of the transformer core is inversely proportional to supply frequency and also the cross-sectional area of the core, the high frequency means that less turns and a smaller core cross-sectional area are required in the transformer. So, the high frequency is lighter, smaller and also low inductance due to less turns.

As for the switches of the bridge, as the switching frequency is high (10,000 Hz is used in the simulation and experiment), and the voltage is not extremely high (the charger is connected to the grid, the boost converter is not taken into consideration in this project), MOSFET(metal-oxide-semiconductor field-effect transistor) is a good choice referring to [41].

## II. SIMULATION BASED ON THE SELECTED TOPOLOGY

### A. FULL BRIDGE TOPOLOGY

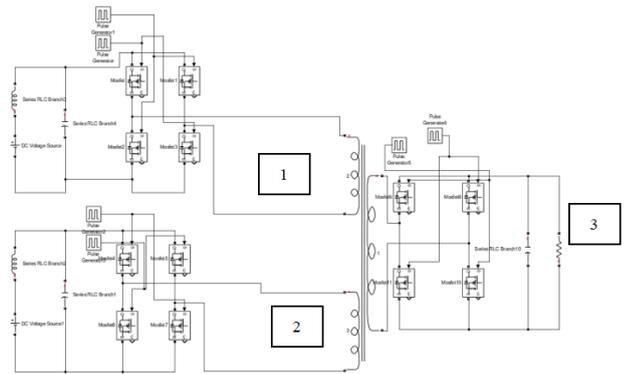

Fig 3.Full bridge topology

Fig. 3 shows the circuit of the tripe full bridge topology. Full bridge means that each bridge contains four switches as an inverter or a rectifier.

- Duty cycle control is applied on the bridge of consuming port:

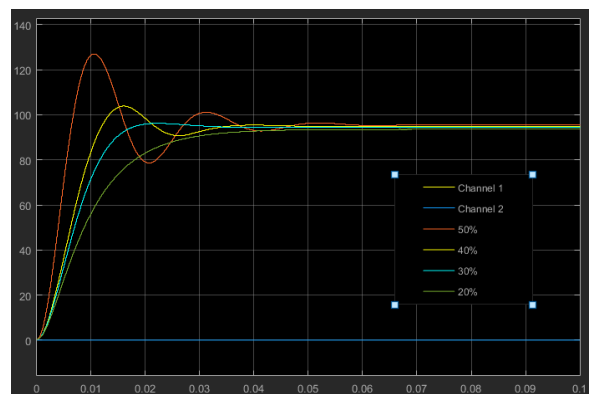

Fig 4.Result of duty cycle control on the receiving port bridge

As the Fig. 4 shown, duty cycle control on the consuming port does not change the steady state output voltage. It only changes their transient state, which may be helpful in the soft start-up.

- Duty cycle control is applied on the bridge of delivering port:

Duty cycle on the delivering port bridge can be used to control the output voltage. The relationship is found to be:

$$V_{out} = D_1 \times V_{in1} \times \frac{n_3}{n_1} \quad (1)$$

Or

$$V_{out} = D_2 \times V_{in2} \times \frac{n_3}{n_2} \quad (2)$$

Where $D_1$ is the duty cycle on the port 1 bridge, $D_2$ is the duty cycle on the port 2 bridge, $n_1$ $n_2$ $n_3$ are the numbers of turns of the transformer for each port.

Fig. 5 shows the result of the sending port duty cycle control. $D_1$ and $D_2$ are changed at the same time and $D_2 = D_2$.

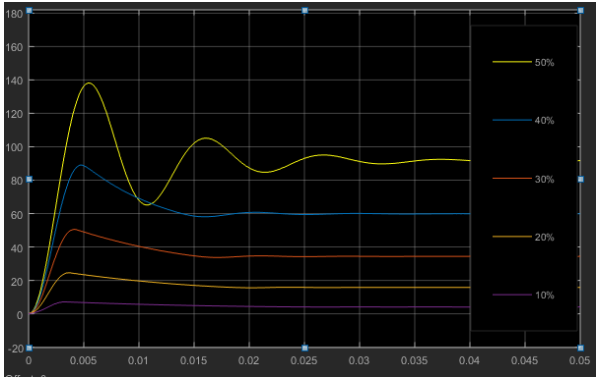

Fig 5. Result of duty cycle control on the sending port bridge

The result of output voltage against duty cycle is plotted as Fig. 6 shown.

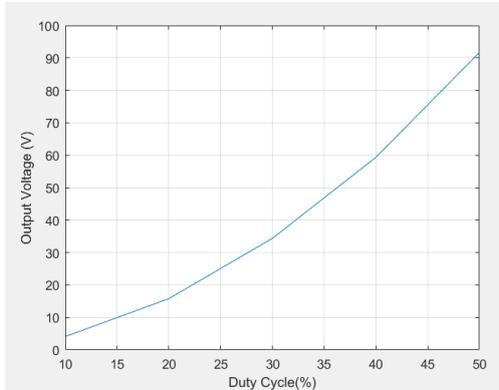

Fig 6. Output voltage vs. duty cycle

The result of output voltage against duty cycle is plotted as Fig 6. shown. An approximately linear relationship can be observed as expected.

- Phase shifting is applied on the bridge of consuming port:

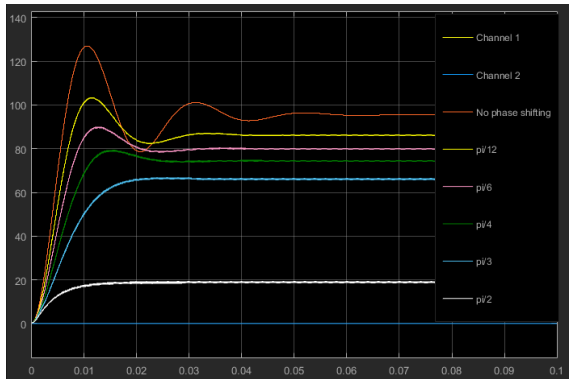

Fig 7. Result of phase shifting

From Fig. 7 we can observe that, as the signal is shifted, the output voltage is decreased. So, phase shifting can be a method to control the output power.

According to [20], if the transformer inductance is taken into consideration, the equivalent circuit is shown in Fig. 8.

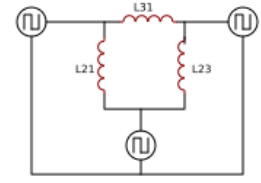

Fig 8. Equivalent circuit of the transformer

Then, the power can be controlled by:

$$P_{21} = S_{21}\varphi_{21}\left(1 - \frac{\varphi_{21}}{\pi}\right) \quad (3)$$

$$P_{31} = S_{31}\varphi_{31}\left(1 - \frac{\varphi_{31}}{\pi}\right) \quad (4)$$

$$P_{23} = S_{23}\varphi_{23}\left(1 - \frac{\varphi_{23}}{\pi}\right) \quad (5)$$

$$S_{21} = \frac{V_2 V_1}{n_2 n_1 \omega L_{21}} \quad (6)$$

$$S_{31} = \frac{V_3 V_1}{n_3 n_1 \omega L_{31}} \quad (7)$$

$$S_{23} = \frac{V_3 V_2}{n_3 n_2 \omega L_{23}} \quad (8)$$

$$\omega = 2\pi f_s \quad (9)$$

And

$$P_1 = P_{21} + P_{31} \quad (10)$$

$$P_2 = P_{21} + P_{23} \quad (11)$$

$$P_3 = P_{23} - P_{31} \quad (12)$$

Where

$f_s$ is the switching frequency.

$\varphi_{21}$ is the phase shifting angle between port 2 and port 1. $\varphi_{31}$ is the phase shifting angle between port 3 and port 1. $\varphi_{23}$ is the phase shifting angle between port 2 and port 3.

$P_1$ is the power delivered by the grid, $P_2$ is the power delivered by the battery and $P_3$ is the power consumed by the load.

*B. HALF BRIDGE TOPOLOGY*

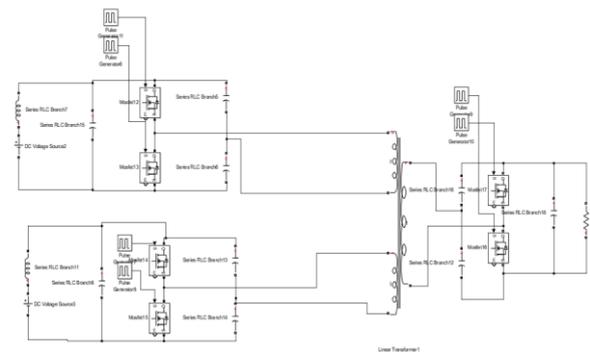

Fig 9. Half bridge topology

Fig. 9 shows the circuit of tripe half bridge topology. The function of the half bridge topology is similar to the full bridge topology. The difference between full bridge topology and half bridge topology is that two switches of the bridge are replaced by two capacitors. So, the circuit of half bridge topology can be cheaper. But the control of the half bridge topology cannot be as complicated as the full bridge topology, which may lead to less efficiency.

As the half bridge topology has the similar structure to the full bridge topology, it shows some same characteristic as the full bridge topology has:

When phase shifting is applied, the output voltage can be control:

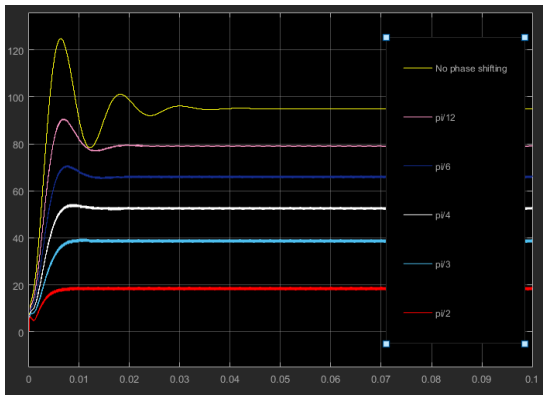
Fig 10.Result of phase shifting

When duty cycle control is applied on the consuming port bridge, the steady output voltage is not changed:

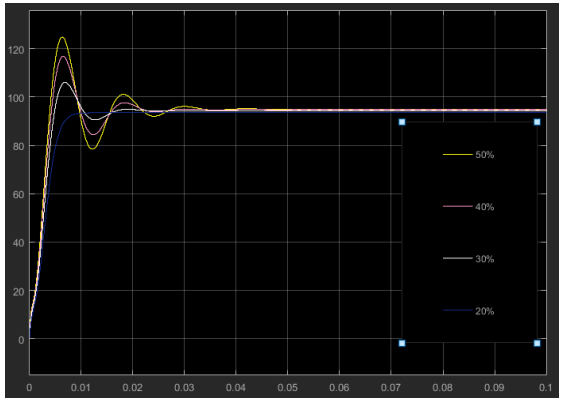
Fig 11.Result of duty cycle control on the receiving port bridge

However, when duty cycle control is applied on the delivering port bridge, different result is obtained:

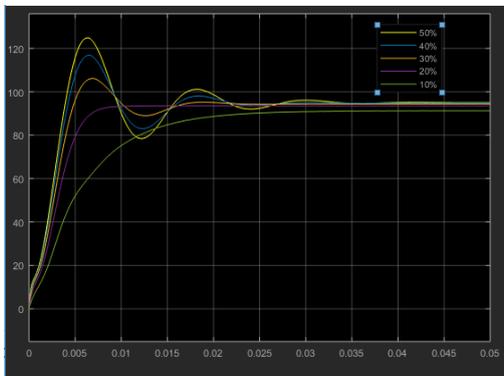
Fig 12.Result of duty cycle control on the sending port bridge

As the Fig. 12 shown, the duty control on the sending port bridge has little effect on the steady output voltage. It is more like the duty cycle control on the consuming port bridge that it only changes the transient state, which is totally different to the full bridge topology.

So, duty cycle control cannot be used in half bridge topology to control the output power.

### C. SIMULATION WITH SINUSOIDAL PWM

There are many types of pulse width modulation (PWM) technique to control the circuit behavior, such as square wave pulse width modulation used in the previous simulation, which is commonly used in power electronic, Sinusoidal PWM and so on. Different technique suits different topology and give different pros and cons.

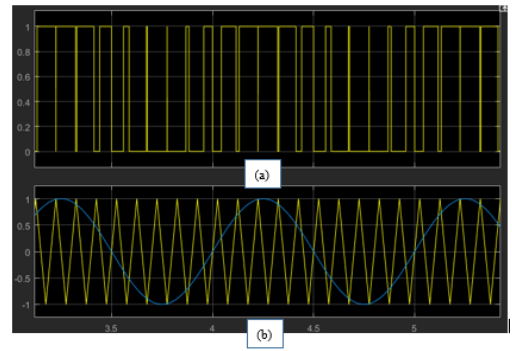
Fig 13.Waveform of SPWM

Sinusoidal pulse width modulation is a modulation technique generated by a triangle wave and sinusoidal wave as the Fig. 13 (b) shown. The switching scheme is shown as follows:

$$V_{out} = +V_{dc}, when\ V_{sine} < V_{triangle} \quad (13)$$
$$V_{out} = -V_{dc}, when\ V_{sine} > V_{triangle} \quad (14)$$

Then the signal of SPWM is generated as the Fig. 13(a).

TABLE 1.Comparison of square wave PWM and Sinusoidal PWM

|  | Square wave PWM | Sinusoidal PWM |
|---|---|---|
| Harmonic distortion | More | Less |
| Control | Duty cycle | amplitude modulation index |
| Switching frequency required | Low | High |

TABLE 1 shows the difference between square wave PWM and sinusoidal PWM. SPWM requires higher switching frequency, but has a less harmonic distortion, which may make a better power transfer through the transformer.

So, a circuit with the combination of square wave PWM and Sinusoidal PWM is developed as the Fig. 14 shown below.

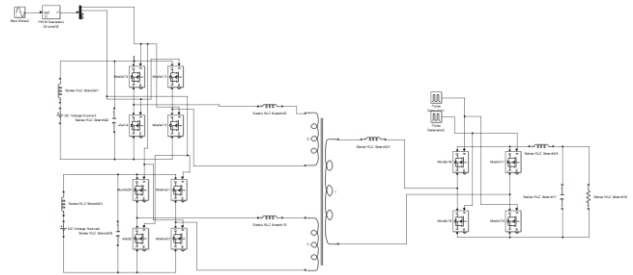
Fig 14.Circuit with combined method

Sinusoidal PWM is applied on the delivering port bridges to make the current through the transformer less distorted, while square wave PWM is applied on the consuming port bridge acting as a rectifier.

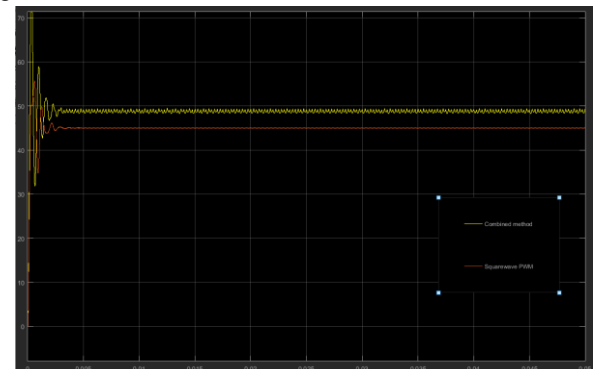
Fig 15.Comparison of the output of two methods

However, when combined method and only square wave PWM control as before at the same circuit with the same size of the filter are compared, the result does not meet the expectation. As shown in the Fig. 15, the voltage with combined method has larger ripples.

Also, phase shifting on the consuming port bridge fails to control the output voltage.

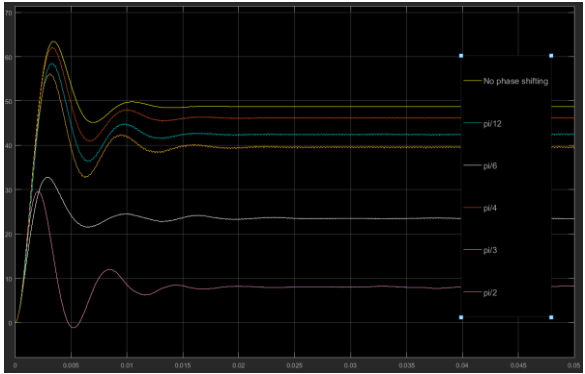
Fig 16.Output of phase shifting

The Fig. 16 shows that the output voltage does not keep being reduced as the phase is shifted. When phase is shifted by pi/4, the output voltage is even higher than it with pi/12 phase shifting. This means that phase shifting control cannot be applied in this condition to control the output power.

There is another way to control the output voltage. The ratio of the amplitudes of reference and carrier signals that is used to generate SPWM signal is called amplitude modulation index:

$$m_a = \frac{V_{m\_\sin}}{V_{m\_tri}} \tag{15}$$

If $m_a \leqq 1$, the amplitude of the fundamental frequency of the output voltage is linearly proportional to $m_a$, which means that we can control the output voltage by changing the amplitude modulation index.

Output voltage can be successfully controlled by changing ma:

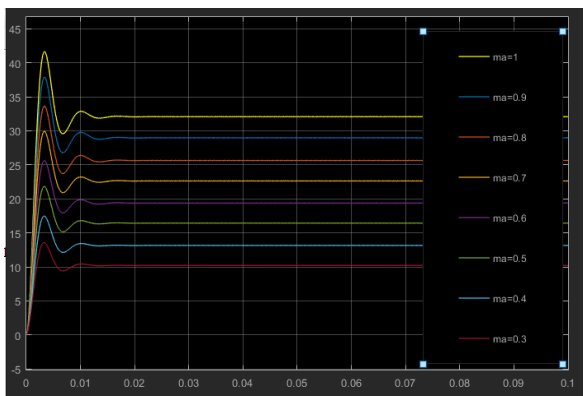
Fig 17.Output of changing amplitude modulation index

The circuit of combined method does not show a better result as expected, which may associate to my raw control method. However, it still has potential to be well developed.

### D. CONNECT TO THE GRID

The previous simulations only use DC voltage source as the power source of the circuit. However, the rapid charging power converter is normally connected to the grid, which means that the three phase AC should be connected as the power source.

So, an AC-DC converter and a capacitor as a DC link can be added to the grid port of the topology as the Fig. 18 shown.

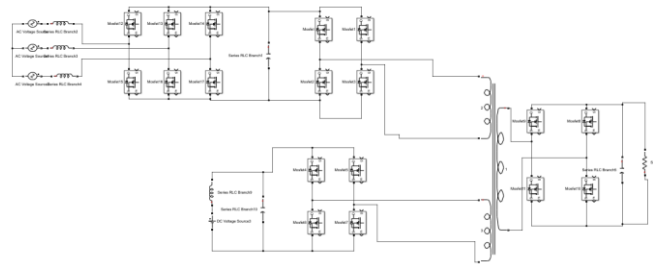
Fig 18.Topology with AC-DC converter

Also, the proposed topology can be extended to a three-phase version as Fig. 19 shown. The required power rating of the components of the circuit can be further reduced. Also, better control can be obtained through all these switches.

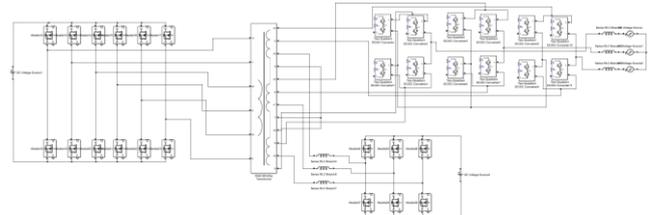
Fig 19.The topology of three phase version

## III. HARDWAVE DESIGN AND EXPERIMENT

As the key point of this topology is to control the switches and make power transferred through the transformer, instead of tripe active bridge, the circuit with one bridge and a transformer is employed

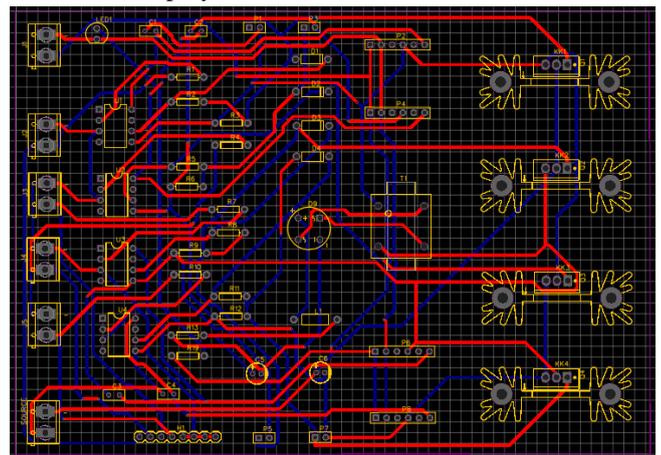
Fig 20.PCB design

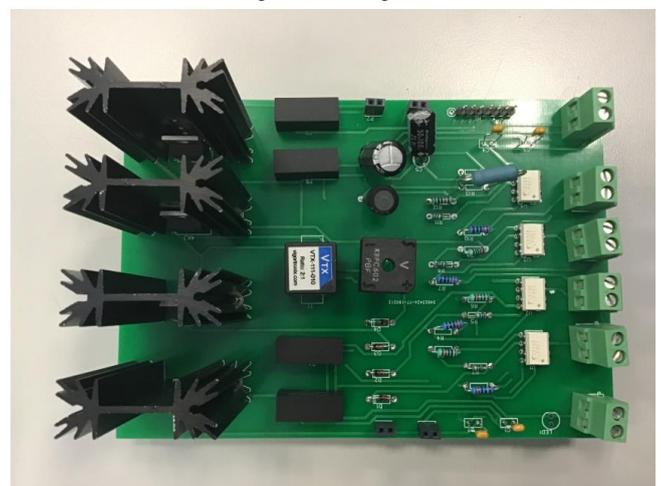
Figure 21.The finished PCB

### A. SIGNAL GENERATION

CY8CKIT-050 PSOC 5LP development kit is used to generate the PWM signal. It enables users to evaluate, develop and prototype high precision analog, low-power and low-voltage applications. It can be programmed by PSOC creator.

PWM signal generated form PSOC development board can be observed by the oscilloscope as Fig. 29 shown. The generated PWM signal has 50% duty cycle and 10 kHz frequency.

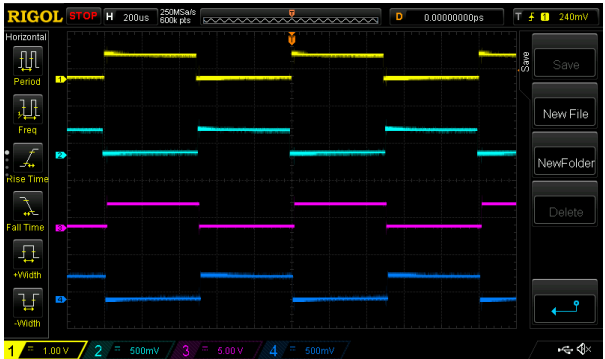
Fig 22.PWM signal generated by PSOC development board

B. WAVEFORMS OF THE CIRCUIT

The input voltage of the gate driver is set to 12V, and the input voltage of the main circuit is set to 15 V. Oscilloscope is used to observe the waveforms of the circuit.

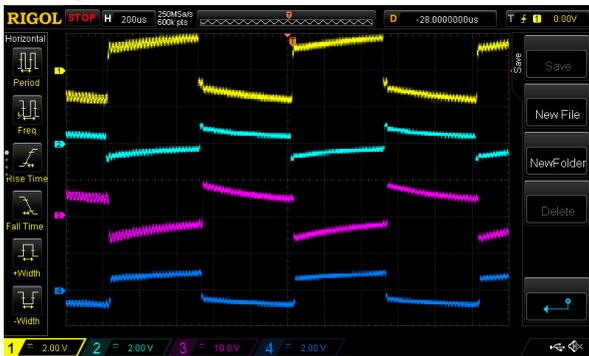
Fig 23.Signal to the MOSFET

Fig. 23 shows the PWM signal apply on the MOSFET. A slight distortion can be observed as the signal is translated through the gate driver.

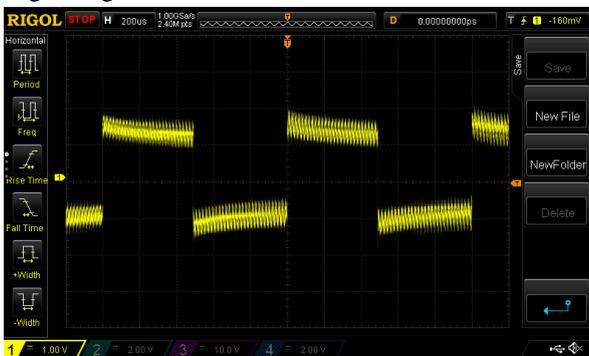
Fig 24.The performance of the MOSFET

The MOSFET works well as Fig. 24 shown.

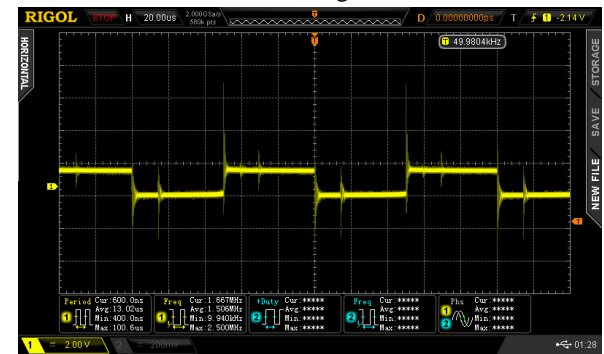
Fig 25.Waveform of the transformer primary side

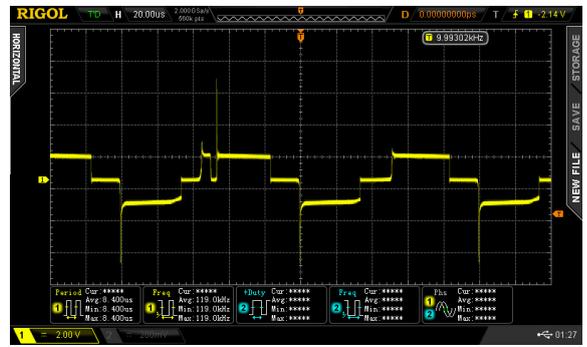
Fig 26.Waveform of the transformer secondary side

Fig. 25 shows the waveform of the transformer primary side while Fig. 26 shows the transformer secondary side. Distortion happens when signal transferred through the transformer. But still, a good output voltage can be maintained as the Fig. 27 shown.

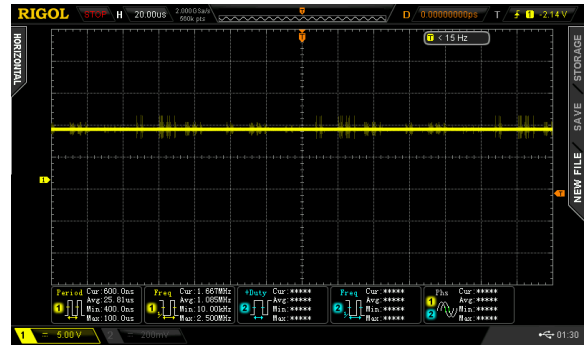
Fig 27.The output waveform

C. DUTY CYCLE CONTROL

Then duty cycle control is applied on the circuit. Duty cycle of the PWM signal can be changed through the PSOC creator.

The result of the duty cycle control is shown as the TABLE 2 below.

TABLE 2. The result of the duty cycle control

| Duty cycle | $V_{OUT}$ (V) | $P_{IN}$ (W) | $P_{OUT}$ (W) | Efficiency |
|---|---|---|---|---|
| 10 | 0.13 | 0.075 | 0.0004225 | 0.005633333 |
| 15 | 0.2 | 0.12 | 0.001 | 0.008333333 |
| 20 | 1.24 | 0.24 | 0.03844 | 0.160166667 |
| 25 | 2.52 | 0.45 | 0.15876 | 0.3528 |
| 30 | 3.6 | 0.63 | 0.324 | 0.514285714 |
| 35 | 4.3 | 0.69 | 0.46225 | 0.669927536 |
| 40 | 5.7 | 1.155 | 0.81225 | 0.703246753 |
| 45 | 6.88 | 1.86 | 1.18336 | 0.636215054 |
| 50 | 8.1 | 2.595 | 1.64025 | 0.632080925 |

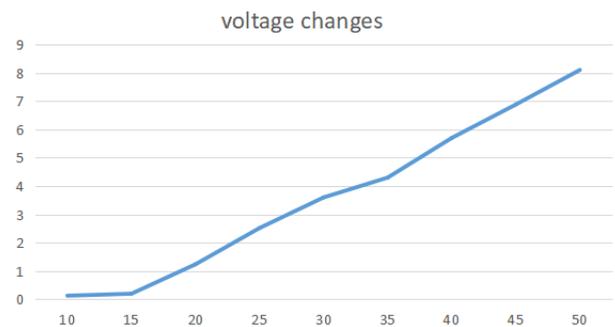
Fig 28.The line chart of voltage changes

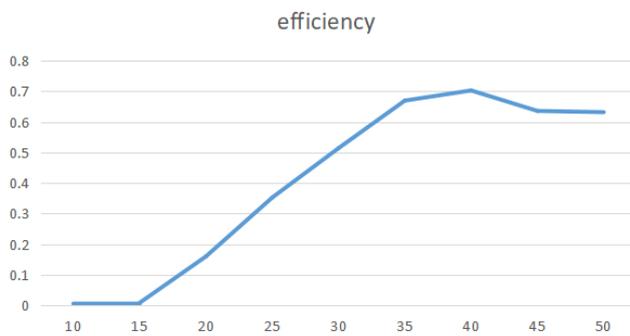

Fig 29. The line chart of the efficiency

The output voltage is successfully controlled as it keep increasing during the increment of the duty cycle.

However, the maximum efficiency appears at around 40% duty cycle.

The efficiency of the circuit is not high. The maximum efficiency is around 70%. The main loss in the circuit happens during the switching as soft switching technique is not applied. Magnetic components losses also contribute the hardware losses.

IV. FUTURE SUGGESTION

The basic function of the rapid charging power converter is achieved in this project. However, soft switching technique is not applied in the simulation and the hardware as the time is limited, which leads to a low efficiency of the converter. Also, the topology is only connected to the DC power supply, which should be connected to an AC power supply.

So, future work can include the soft switching technique, and also the control method and simulation of the topology connected to the grid.

V. CONCLUSION

The basic multiport rapid charging topology is discussed in this report. Some related simulations are done. And the hardware circuit with one bridge and one transformer is designed and tested. The idea of SPWM control is proposed in this project, and it still need to be improved.